\documentclass[a4paper]{jpconf}
\usepackage{amsmath}
\usepackage{graphicx}

\begin{document}

\title{Secular momentum transport by gravitational waves from spinning
compact binaries}
\author{Zolt\'{a}n Keresztes$^{1,2\star }$, Bal\'{a}zs Mik\'{o}czi$^{3\dag }$%
, L\'{a}szl\'{o} \'{A}. Gergely$^{1,2,4\ddag }$, M\'{a}ty\'{a}s Vas\'{u}th$^{3%
\P } $}

\begin{abstract}
We present a closed system of coupled first order differential equations
governing the secular linear momentum loss of a compact binary due to
emitted gravitational waves, with the leading order relativistic and
spin-orbit perturbations included. In order to close the system, the secular
evolution equations of the linear momentum derived from the dissipative
dynamics are supplemented with the secular evolutions of the coupled angular
variables, as derived from the conservative dynamics.
\end{abstract}

\address{$^{1}$ Department of Theoretical Physics, University of Szeged, Tisza Lajos
krt 84-86, Szeged 6720, Hungary\\
$^{2}$ Department of Experimental Physics, University of Szeged, D\'{o}m t\'{e}r 9, Szeged 6720, Hungary\\
$^{3}$ KFKI Research Institute for Particle and Nuclear Physics, Budapest
114, P.O.Box 49, H-1525 Hungary\\
$^{4}$ Institute for Advanced Study, Collegium Budapest, Szenth\'aroms\'ag u 2, Budapest 1014, Hungary}

\ead{$^{\ast}$zkeresztes@titan.physx.u-szeged.hu\quad $^{\dag}$mikoczi@rmki.kfki.hu
\quad $^{\ddag }$gergely@physx.u-szeged.hu\quad $^{\P}$vasuth@rmki.kfki.hu}

\section{Introduction}

The inspiral of compact objects in binary systems is driven by the emitted
gravitational radiation which carries energy, angular and linear momenta
away from the source. Global conservation of these quantities implies a
radiative orbital evolution, including a possible recoil of the system. The
energy and magnitude of orbital angular momentum of the binary determine the
quasi-Keplerian orbit in the post-Newtonian (PN) regime. These quantities
are known to high accuracy for binaries on eccentric orbits, with PN,
spin-orbit (SO), spin-spin (SS), mass quadrupole - mass dipole (QM) and
magnetic dipole - magnetic dipole (DD) coupling terms \cite{Kepler}.
Moreover, the orientation of the (quasi-precessing) plan of motion, defined
by the direction of the Newtonian orbital angular momentum, is known to high
accuracy with all the above-mentioned contributions.

Due to asymmetries in the configuration of the source the radiation is often
emitted anisotropically. The linear momentum loss of the binary leads to the
recoil of the center of mass in the opposite direction, which in extreme
situations results the kick-off of the binary from its host galaxy. This
effect, however, vanishes for equal mass binaries due to the fact that the
leading order relativistic contribution to the linear momentum loss scales
with the mass difference $\delta m=m_{1}-m_{2}$ of the orbiting bodies.

The gravitational recoil of a binary system and the classical linear
momentum loss of the final black hole were discussed in \cite%
{Peres,Bekenstein} with the inclusion of the lowest multipoles needed for
the computation of momentum ejection. The first quasi-Keplerian analytic
studies were given in \cite{Fitchett}, where the linear momentum flux of
gravitational waves from a binary system of two point masses in Keplerian
orbit were calculated. 1\ PN corrections to the gravitational recoil were
discussed in \cite{Wiseman}, while 2PN recoil effects for binaries on
quasicircular orbits in \cite{Blanchet}. Several numerical estimates for the
kick velocity of non-spinning binaries were given, e.g. in \cite%
{Baker,HHind,GonzalezNS}, with the maximum recoil in this case found as $%
\sim 175$ km/s \cite{Gonzalez}. Recently it has been shown \cite{TBW}, that
the ringdown phase acts as an anti-kick as compared to the inspiral and
plunge, the global result being consistent with the numerical estimates.

The rotation of the components adds however additional structure to the
emitted radiation and recoil. Spin contributions to the linear momentum loss
are analyzed in \cite{Kidder}, and more recently in \cite{SchBuon}. The SO
contributions scale with the magnitude of spins which could be high for
galactic black holes. Numerical analyses indicate that significant
gravitational recoil can be obtained in spinning binaries \cite%
{Herrmann,Koppitz,Campanelli,Schnittman1,Lousto} even with a high degree of
symmetry in the configuration, i.e. for equal-mass binaries with antialigned
initial spins in the orbital plane \cite{Brugmann}. In a detailed review 
\cite{Racine}, Racine, Buonanno and Kidder have computed the instantaneous
linear momentum flux emitted by spinning binaries at 2PN order with the
inclusion of the next-to-leading order SO, SO tail and SS terms. Moreover,
the recoil velocity as a function of the orbital frequency was given for
quasicircular orbits.

In the present work we give the previously unknown \textit{secular
expressions} for the linear momentum loss of an inspiralling binary system,
with the inclusion of leading order relativistic and SO contributions. The
analytic approach presented here is suitable for characterize analytically
the dependence of the recoil on binary parameters.

Section 2 contains elements of the conservative dynamics with the inclusion
of first post-Newtonian and spin-orbit contributions. At the end of the
section the secular conservative evolution of the relevant angle variables
is given. In Section 3 we start from the expressions of the SO contributions
to the instantaneous linear momentum loss given in \cite{Kidder}, then we
derive the dissipative secular linear momentum evolution. The two sets of
secular evolutions couple to a closed system of first order differential
equations.

The secular evolutions are derived as follows: first we rewrite the
instantaneous evolutions in terms of the radial parametrization of
quasi-Keplerian orbits \cite{Kepler}. As a result, all the radial functional
dependencies are expressed in terms of a single variable $\chi $, the
generalized true anomaly. Averaging these expressions over a radial period
becomes particularly straightforward in terms of the complex version of $%
\chi $, which renders the problem to the computation of residues in the
origin of the complex parameter plane \cite{Param,Gergely}. This procedure
smears out short timescale effects and we obtain a simpler dynamics,
suitable for monitoring the secular changes. The procedure is similar to our
previous computations of the SO-induced secular changes of the energy,
magnitude of orbital angular momentum and relative angles among the orbital
angular momentum and spins \cite{GPV3}. An additional difficulty in the
present computation arises from the vectorial character of the linear
momentum, and the subsequent system of coupled differential equations.

Our description is generic, being valid for generic (non-circular,
non-spherical) orbits.

\textit{Notation.} For any vector $\mathbf{V}$ its magnitude is denoted as $%
V $ and its direction (a unit vector) as $\mathbf{\hat{V}}$.

\section{Elements of the conservative dynamics}

In this section we derive useful elements of the conservative dynamics,
originating from the SO coupling. We derive the result in terms of:

(a) the physical parameters of the binary: total mass $m=m_{1}+m_{2}$, mass
ratio ${\nu =m}_{2}/m_{1}\leq 1$, symmetric mass ratio $\eta =\mu /m$ (where 
$\mu =m_{1}m_{2}/m$ is the reduced mass), dimensionless spin parameters $%
\chi _{i}=\left( G/c\right) ^{-1}S_{i}/m_{i}^{2}$

(b) dynamical constants of motion: the energy $E$, the magnitude $L$ of the
orbital angular momentum $\mathbf{L}$ and $A=\left( G^{2}m^{2}\mu ^{2}+{%
2EL^{2}/\mu }\right) ^{1/2}$ (this would be the length of the
Laplace-Runge-Lenz vector of the Keplerian motion characterized by $E$ and $%
\mathbf{L}$)

(c) angular variables related to the orbital momenta: the relative angles of
spins among themselves $\gamma $ and with the orbital angular momentum $%
\kappa _{i}$, the polar angles $\psi _{i}$ of the spins in the plane of
motion (measured from the intersection $\mathbf{\hat{l}}$ of the plane of
motion with the plane perpendicular to the total angular momentum $\mathbf{%
J=L+S}$, where $\mathbf{S=S}_{\mathbf{1}}+\mathbf{S}_{\mathbf{2}}$), and

(d) angular variables characterizing the orbit: the inclination $\alpha $ of
the orbital plane with respect to the plane perpendicular to $\mathbf{J}$,
the angle $\phi _{n}$ between $\mathbf{\hat{l}}$ and an inertial axis $%
\mathbf{\hat{x}}$ taken in the plane perpendicular to $\mathbf{J}$, finally
the angle $\psi _{p}$ span by the periastron $\mathbf{\hat{A}}_{\mathbf{N}}$
and $\mathbf{\hat{l}}$ (see Fig 1). These three angles will be referred
occasionally as Euler angles, as three consecutive rotations with $\phi
_{n},~\alpha $ and $\psi _{p}$ about the axes $z,~x$ and again $z$ transform
from an inertial system with $\mathbf{\hat{J}}$ as the \thinspace $z\,$-axis
to the system with the $x$-axis pointing towards the periastron and the $y$%
-axis in the plane of motion. 
\begin{figure}[th]
\begin{center}
\includegraphics[height=10cm]{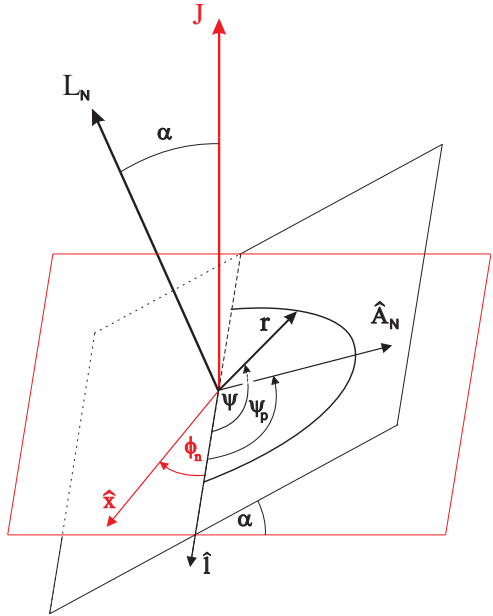}
\end{center}
\caption{The angular variables.}
\label{fig1}
\end{figure}

\subsection{Lagrangian dynamics}

The Lagrangian characterizing the dynamics to 1.5 PN orders (tail terms are
omitted) is

\begin{equation}
\mathcal{L}=\mathcal{L}_{N}+\mathcal{L}_{PN}+\mathcal{L}_{SO}\ ,
\end{equation}%
with the contributions%
\begin{eqnarray}
\mathcal{L}_{N} &=&\frac{\mu \mathbf{v}^{2}}{2}+\frac{Gm\mu }{r}\ ,  \notag
\\
\mathcal{L}_{PN} &=&\frac{1}{8c^{2}}\left( 1-3\eta \right) \mu v^{4}+\!\frac{%
Gm\mu }{2rc^{2}}\!\left[ \!\left( 3\!+\!\eta \right) v^{2}\!+\!\eta \dot{r}%
^{2}\!-\!\frac{Gm}{r}\!\right] \ ,  \notag \\
\mathcal{L}_{SO} &=&\frac{G\mu }{2c^{2}r^{3}}\mathbf{v}\cdot \lbrack \mathbf{%
r}\times (4\mathbf{S}+3{\mbox{\boldmath $\sigma$}})]\ .
\end{eqnarray}%
Here ${\mbox{\boldmath
$\sigma$}=\nu }\mathbf{S}_{\mathbf{1}}+{\nu }^{-1}\mathbf{S}_{\mathbf{2}}$
is the mass-weighted spin vector. The SO part of the Lagrangian applies to
the Newton-Wigner spin supplementary condition and was given first in Ref. 
\cite{Kepler}. The Lagrangian gives%
\begin{equation}
\mathbf{a}=\mathbf{a}_{\mathbf{N}}+\mathbf{a}_{\mathbf{PN}}+\mathbf{a}_{%
\mathbf{SO}}\ ,  \label{acc}
\end{equation}%
with 
\begin{eqnarray}
\mathbf{a}_{\mathbf{N}} &=&-\frac{Gm}{r^{3}}\mathbf{r}\ , \\
\mathbf{a}_{\mathbf{PN}} &=&-\frac{Gm}{c^{2}r^{3}}\left\{ \mathbf{r}\left[
\left( 1+3\eta \right) \mathbf{v}^{2}-2\left( 2+\eta \right) \frac{Gm}{r}-%
\frac{3}{2}\eta \dot{r}^{2}\right] -2\left( 2-\eta \right) r\dot{r}\mathbf{v}%
\right\} \ , \\
\mathbf{a}_{\mathbf{SO}} &=&\frac{G}{c^{2}r^{3}}\left[ \frac{3}{2r^{2}}%
\mathbf{r}\left[ \left( \mathbf{r}\times \mathbf{v}\right) \cdot \left( 4%
\mathbf{S}+3{\mbox{\boldmath $\sigma$}}\right) \right] +\frac{3}{2}\frac{%
\dot{r}}{r}\mathbf{r}\times \left( 4\mathbf{S}+3{\mbox{\boldmath $\sigma$}}%
\right) -\mathbf{v}\times \left( 4\mathbf{S}+3{\mbox{\boldmath $\sigma$}}%
\right) \right] \ .
\end{eqnarray}

\subsection{Constants of motion}

The constants of motion are the energy $E$ and the total angular momentum $%
\mathbf{J}$. As the total orbital angular momentum undergoes a pure
precessional motion \cite{BOC1,BOC2}, its magnitude $L$ is also a conserved
quantity.

The total orbital angular momentum $\mathbf{L=r\times }\left( \partial 
\mathcal{L}/\partial \mathbf{v}\right) $ can be decomposed as 
\begin{equation}
\mathbf{L}=\mathbf{L}_{\mathbf{N}}+\mathbf{L}_{\mathbf{PN}}+\mathbf{L}_{%
\mathbf{SO}}\ ,  \label{Ltot}
\end{equation}%
with 
\begin{eqnarray}
\mathbf{L_{N}} &=&\mu \mathbf{r}\times \mathbf{v}\ ,  \label{LN} \\
\mathbf{L_{PN}} &=&\mathbf{L_{N}}\left[ \frac{\left( 1-3\eta \right) }{2}%
\frac{\mathbf{v}^{2}}{c^{2}}+\left( 3\!+\!\eta \right) \!\frac{Gm}{c^{2}r}%
\right] \ , \\
\mathbf{L_{SO}} &=&\frac{G\mu }{2c^{2}r^{3}}\mathbf{r}\times \left[ \mathbf{r%
}\times \left( 4\mathbf{S}+3{\mbox{\boldmath $\sigma$}}\right) \right] \ .
\end{eqnarray}%
We note here the useful approximate relation%
\begin{equation}
L^{2}={L}_{N}^{2}+\lambda _{PN}+\lambda _{SO}\ ,  \label{L2}
\end{equation}%
where%
\begin{eqnarray}
\lambda _{PN} &\equiv &2\mathbf{L}_{\mathbf{N}}\cdot \mathbf{L}_{\mathbf{PN}%
}=2{L}^{2}\left[ \left( 1-3\eta \right) \frac{E}{c^{2}}+2\left( 2-\eta
\right) \!\frac{Gm\mu }{c^{2}r}\right] \ , \\
\lambda _{SO} &\equiv &2\mathbf{L}_{\mathbf{N}}\cdot \mathbf{L}_{\mathbf{SO}%
}=-\frac{G\mu L}{c^{2}r}\sum\limits_{i=1,j\neq i}^{2}\!\frac{4m_{i}+3m_{j}}{%
m_{i}}S_{i}\cos \kappa _{i}\ .
\end{eqnarray}

\subsection{The evolutions of $\protect\psi$ and $\protect\phi _{n}$}

The polar and azimuthal angles $\theta ,\varphi $ of the reduced mass
particle in the inertial system with $\mathbf{\hat{J}}$ as the \thinspace $%
z\,$-axis can be related to the Euler angles ($\phi _{n},~\alpha ,~\psi $)
characterizing a non-inertial reference system with the $x$-axis at the
location of the reduced mass particle and the $y$-axis in the plane of
motion \cite{GPV1}: 
\begin{eqnarray}
\sin \theta \cos \varphi  &=&\cos \alpha \sin \phi _{n}\sin \psi +\cos \phi
_{n}\cos \psi \ ,  \label{Euler1} \\
\sin \theta \sin \varphi  &=&\cos \alpha \cos \phi _{n}\sin \psi -\sin \phi
_{n}\cos \psi \ ,  \label{Euler2} \\
\cos \theta  &=&\sin \alpha \sin \psi \ .  \label{Euler3}
\end{eqnarray}%
Tedious but straightforward algebra leads to%
\begin{eqnarray}
\dot{\psi} &=&\cos \alpha \ \dot{\varphi}-\frac{\sin \alpha \cos \psi }{\sin
\theta }\ \dot{\theta}+\cos \alpha \ \dot{\phi}_{n}\ ,  \notag \\
\dot{\phi}_{n} &=&-\dot{\varphi}-\frac{\cot \alpha }{\sin \theta \cos \psi }%
\ \dot{\theta}+\frac{\tan \psi }{\sin ^{2}\alpha }\frac{d}{dt}\left( \cos
\alpha \right) ~\ .  \label{angles_t}
\end{eqnarray}%
The Newtonian orbital angular momentum, expressed in terms of ($\theta $, $%
\varphi $) gives%
\begin{equation}
L_{N}^{2}={\mu ^{2}}r^{4}(\dot{\theta}^{2}+\sin ^{2}\theta \ \dot{\varphi}%
^{2})\ ,\quad \left( \mathbf{L_{N}}\right) _{z}={\mu }r^{2}\sin ^{2}\theta \ 
\dot{\varphi}\ .
\end{equation}%
By employing Eqs. (\ref{L2}), (\ref{Euler3}) and $\left( \mathbf{L_{N}}%
\right) _{z}=L_{N}\cos \alpha $ we obtain the evolutions:%
\begin{eqnarray}
\sin \theta ~\dot{\theta} &=&-\frac{L}{{\mu }r^{2}}\sin \alpha \cos \psi
\left( 1-\frac{\lambda _{PN}+\lambda _{SO}}{2{L}^{2}}\right) \ ,
\label{thetadotsquare} \\
\sin ^{2}\theta ~\dot{\varphi} &=&\frac{L}{{\mu }r^{2}}\cos \alpha \left( 1-%
\frac{\lambda _{PN}+\lambda _{SO}}{2L^{2}}\right) \ .
\end{eqnarray}%
The minus sign in the first equation was chosen after taking the square root
in such a way that when we insert these equations into Eqs. (\ref{angles_t}%
), we obtain 
\begin{eqnarray}
\dot{\psi} &=&\frac{L}{{\mu }r^{2}}\left( 1-\frac{\lambda _{PN}+\lambda _{SO}%
}{2{L}^{2}}\right) +\dot{\phi}_{n}\cos \alpha \ \ ,  \label{psidot} \\
\dot{\phi}_{n} &=&\frac{\tan \psi }{\sin ^{2}\alpha }\frac{d}{dt}\left( \cos
\alpha \right) \ ,  \label{phindot}
\end{eqnarray}%
which reduce to the correct equations $\dot{\psi}=L/\mu r^{2}$ and $\phi
_{n}=$const at Newtonian order.

Provided that we can obtain $\dot{\alpha}$ by a complementary method, the
evolutions of $\psi $ and $\phi _{n}$ can be given explicitly. This will be
done in the following subsection.

\subsection{The evolution of the orbital inclination}

From $\cos \alpha =\mathbf{\hat{J}}\cdot \mathbf{\hat{L}}_{\mathbf{N}}$, as $%
\mathbf{J}$ is conserved, we get%
\begin{equation}
\frac{d}{dt}\left( \cos \alpha \right) =\mathbf{\hat{J}}\cdot \frac{d}{dt}%
\left( \mathbf{\hat{L}}_{\mathbf{N}}\right) \ .  \label{alphadot}
\end{equation}%
Employing Eq. (\ref{acc}) gives 
\begin{equation}
\mathbf{\dot{L}}_{\mathbf{N}}=\mu \mathbf{r}\times \mathbf{a=}\ \frac{2Gm}{%
c^{2}r^{2}}\left( 2-\eta \right) \dot{r}\mathbf{L}_{\mathbf{N}}+\frac{G\mu }{%
c^{2}r^{3}}\left\{ \left( \frac{3}{2}\frac{\dot{r}}{r}\ \mathbf{r}-\mathbf{v}%
\right) \left[ \mathbf{r}\cdot \left( 4\mathbf{S}+3{\mbox{\boldmath $\sigma$}%
}\right) \right] -\frac{r\dot{r}}{2}\left( 4\mathbf{S}+3{%
\mbox{\boldmath
$\sigma$}}\right) \right\} \ .
\end{equation}%
From here we can derive the time derivative of $\mathbf{\hat{L}}_{\mathbf{N}}
$. We then find that the leading order evolution of $\alpha $ is of order $%
S^{2}/JL$:%
\begin{eqnarray}
\frac{d}{dt}\left( \cos \alpha \right)  &=&\frac{G\mu \dot{r}}{2c^{2}JLr^{2}}%
\left[ \left( 4+3\nu \right) S_{1}\cos \kappa _{1}+\left( 4+3\nu
^{-1}\right) S_{2}\cos \kappa _{2}\right] \left( S_{1}\cos \kappa
_{1}+S_{2}\cos \kappa _{2}\right)   \notag \\
&&-\frac{G\mu \dot{r}}{2c^{2}JLr^{2}}\left\{ \left( 4+3\nu \right) S_{1}^{2}+%
\left[ 8+3\left( \nu +\nu ^{-1}\right) \right] S_{1}S_{2}\cos \gamma +\left(
4+3\nu ^{-1}\right) S_{2}^{2}\right\}   \notag \\
&&+\frac{G\mu }{c^{2}JLr^{3}}\left[ \left( 4+3\nu \right) \mathbf{r\cdot S}_{%
\mathbf{1}}+\left( 4+3\nu ^{-1}\right) \mathbf{r\cdot S}_{\mathbf{2}}\right]
\left( \frac{3}{2}\frac{\dot{r}}{r}\ \mathbf{r}-\mathbf{v}\right) \mathbf{%
\cdot }\left( \mathbf{S}_{\mathbf{1}}+\mathbf{S}_{\mathbf{2}}\right) \ .
\label{cosalphadot}
\end{eqnarray}%
(We have employed $L_{N}=L$, valid to leading order accuracy.) The scalar
products can be rewritten in terms of the angular variables as%
\begin{eqnarray}
\mathbf{r\cdot S_{i}} &=&rS_{i}\sin \kappa _{i}\cos (\psi _{p}+\chi -\psi
_{i})\ ,  \notag \\
\mathbf{v\cdot S_{i}} &=&\dot{r}S_{i}\sin \kappa _{i}\cos (\psi _{p}+\chi
-\psi _{i})-\frac{LS_{i}}{\mu r}\sin \kappa _{i}\sin (\psi _{p}+\chi -\psi
_{i})\ ,
\end{eqnarray}%
In Eq. (\ref{cosalphadot}), as all terms are of $S^{2}/J$ order, it is
allowed to employ the Newtonian true anomaly parametrization of the orbit:%
\begin{equation}
r=\frac{L^{2}}{\mu (Gm\mu +A\cos \chi )}\ ,\quad \dot{r}={\frac{A}{L}}\sin
\chi \ .  \label{newt}
\end{equation}%
With these expressions, all the desired angular evolutions can be given
explicitly as functions of $\chi $ (which, to Newtonian order is $\psi -\psi
_{p}$, see Figure 1).

\subsection{The secular evolution of the Euler angles}

We conclude this section by giving the secular evolutions of the Euler
angles. The secular evolution of any function $f\left( \chi \right) $ is
defined as $<\dot{f}>=T^{-1}\int_{0}^{2\pi }\dot{f}\left( \chi \right) $ $%
\dot{\chi}^{-1}$ $d\chi $. As the instantaneous angular evolutions (\ref%
{phindot}) and (\ref{cosalphadot}) have no Newtonian contributions, we can
employ the Newtonian period of the orbital motion $T=2\pi Gm\left[ \mu /(-2E)%
\right] ^{3/2}$ in calculating their orbital average, obtaining:%
\begin{eqnarray}
\sin \alpha \left\langle \dot{\alpha}\right\rangle &=&\frac{3G\left( -2E\mu
\right) ^{3/2}}{2c^{2}JL^{3}}\left( \nu -\nu ^{-1}\right) S_{1}S_{2}\sin
\kappa _{1}\sin \kappa _{2}\sin \Delta \psi \ ,  \notag \\
\sin ^{2}\alpha \ \left\langle \dot{\phi}_{n}\right\rangle &=&\frac{G\left(
-2E\mu \right) ^{3/2}}{2c^{2}JL^{3}}\Biggl[\sum\limits_{i=1}^{2}\left(
4+3\nu ^{3-2i}\right) S_{i}^{2}\sin \kappa _{i}^{2}\cos 2\psi _{i}  \notag \\
&&+S_{1}S_{2}\left[ 8+3\left( \nu +\nu ^{-1}\right) \right] \sin \kappa
_{1}\sin \kappa _{2}\cos 2\overline{\psi }\Biggr]\ .  \label{angledotave}
\end{eqnarray}%
We introduced here the shorthand notations $\Delta \psi =\psi _{2}-\psi _{1}$
and $\overline{\psi }=\left( \psi _{1}+\psi _{2}\right) /2$.

We also compute the change in the angle $\psi $ over a radial period by the
above indicated method, as $\Delta \psi =\int_{0}^{2\pi }\dot{\psi}\left(
\chi \right) $ $\dot{\chi}^{-1}$ $d\chi .$ By employing Eq. (\ref{psidot})
and the relevant terms\footnote{%
The contributions $\left( dt/d\chi \right) _{PN}$ and $\left( dt/d\chi
\right) _{SO}$ are corrected by a global sign and the factor $\mu /L$,
respectively.} form Eq. (18) of Ref. \cite{Kepler} we get:%
\begin{eqnarray}
\frac{dt}{d\chi } &=& \left( \frac{dt}{d\chi }\right) _{N}\!+\left( \frac{dt%
}{d\chi }\right) _{PN}+\left( \frac{dt}{d\chi }\right) _{SO}\ , \\
\left( \frac{dt}{d\chi }\right) _{N} &=& \frac{\mu r^{2}}{L}\ ,  \notag \\
\left( \frac{dt}{d\chi }\right) _{PN} &=& -\frac{\mu r^{2}}{2c^{2}L^{3}}%
\left[ \left( \eta -13\right) G^{2}m^{2}\mu ^{2}+\left( 3\eta -1\right)
A^{2}+(3\eta -8)Gm\mu A\cos \chi \right] \ ,  \notag \\
\left( \frac{dt}{d\chi }\right) _{SO} &=& -\frac{G\mu ^{3}r^{2}}{2c^{2}L^{4}}%
\left( 3Gm\mu +A\cos \chi \right) B_{S}\ ,
\end{eqnarray}%
with 
\begin{equation}
B_{S}=\left( 4+3\nu \right) S_{1}\cos \kappa _{1}+\left( 4+3\nu ^{-1}\right)
S_{2}\cos \kappa _{2}\ .
\end{equation}%
The leading order term gives $\int_{0}^{2\pi }\left( L/{\mu }r^{2}\right) $ $%
\left( dt/d\chi \right) _{N}$ $d\chi =2\pi .$ Defining $\left\langle \dot{%
\psi}_{p}\right\rangle \equiv \left( \Delta \psi -2\pi \right) /T$ we get
the secular periastron precession rate as 
\begin{equation}
\left\langle \dot{\psi}_{p}\right\rangle =\frac{Gm\left( -2E\right)
^{3/2}\mu ^{1/2}}{c^{2}L^{2}}\left( 3-\eta \frac{B_{S}}{L}\right) +\cos
\alpha \ \left\langle \dot{\phi}_{n}\right\rangle \ .  \label{psipdotave}
\end{equation}%
Eqs. (\ref{angledotave}) and (\ref{psipdotave}) together with the evolution
of the angles $\kappa _{i}$ given by the first two Eqs. (2.17) of \cite{GPV3}
and the projections of $\mathbf{J=L+S}_{\mathbf{1}}+\mathbf{S}_{\mathbf{2}}$%
, determining $\psi _{i}$ in terms of the other variables form a closed
system of differential equations for the conservative evolution of the
angular variables.

\section{Elements of the dissipative dynamics}

The instantaneous loss in the linear momentum due to the gravitational
radiation can be expressed as $\mathbf{\dot{P}}_{i}=-\{{\frac{2}{63}}\overset%
{(4)}{I_{ijk}}\overset{(3)}{I_{jk}}+{\frac{16}{45}}\epsilon _{ijk}\overset{%
(3)}{I_{jl}}\overset{(3)}{J_{kl}}\}$, 
where $I_{jk}$, $I_{ijk}$ and $J_{kl}$ are the mass quadrupole, mass
octupole and current quadrupole moments \cite{Kidder}. The secular loss in
the linear momentum is:%
\begin{equation}
\left\langle \mathbf{\dot{P}}\right\rangle =\left\langle \mathbf{\dot{P}}%
\right\rangle _{N}+\left\langle \mathbf{\dot{P}}\right\rangle _{SO}\ ,
\label{Pdotave}
\end{equation}%
with the leading order radiative (N) and leading order radiative SO
contributions given by 
\begin{eqnarray}
\left\langle \mathbf{\dot{P}}\right\rangle _{N}\!\!\! &=&\!\!\!-\delta m%
\frac{G^{2}\mu ^{3}A}{30c^{7}L^{8}}\left( \frac{-2E}{\mu }\right) ^{3/2}%
\mathcal{A}_{N}\mathbf{\ U}\ , \\
\left\langle \mathbf{\dot{P}}\right\rangle _{SO}\!\!\! &=&\!\!\!\frac{%
G^{2}\mu ^{4}A}{30c^{7}L^{9}}\left( \frac{-2E}{\mu }\right) ^{3/2}  \notag \\
&&\times \sum\limits_{i=1}^{2}\left( -1\right) ^{i-1}\left( 1+\nu
^{3-2i}\right) S_{i}\left[ \mathcal{A}_{SO}\ \mathbf{U}\cos \kappa _{i}-%
\mathcal{B}_{SO}\mathbf{V}\sin \kappa _{i}\sin \left( \psi _{i}-\psi
_{p}\right) \right] .
\end{eqnarray}%
Here we have denoted 
\begin{equation}
\mathbf{U=}\left( 
\begin{array}{c}
\cos \alpha \sin \phi _{n}\cos \psi _{p}-\cos \phi _{n}\sin \psi _{p} \\ 
\cos \alpha \cos \phi _{n}\cos \psi _{p}+\sin \phi _{n}\sin \psi _{p} \\ 
\sin \alpha \cos \psi _{p}%
\end{array}%
\right) \ ,\quad \mathbf{V=}\left( 
\begin{array}{c}
-\sin \alpha \sin \phi _{n} \\ 
-\sin \alpha \cos \phi _{n} \\ 
\cos \alpha%
\end{array}%
\right) ~,
\end{equation}%
and%
\begin{eqnarray}
\mathcal{A}_{N} &=&148E^{2}L^{4}+1060G^{2}m^{2}\mu
^{3}EL^{2}+805G^{4}m^{4}\mu ^{6}\ , \\
\mathcal{A}_{SO} &=&48E^{2}L^{4}+360G^{2}m^{2}\mu
^{3}EL^{2}+280G^{4}m^{4}\mu ^{6}\ , \\
\mathcal{B}_{SO} &=&108E^{2}L^{4}+780G^{2}m^{2}\mu
^{3}EL^{2}+595G^{4}m^{4}\mu ^{6}\ .
\end{eqnarray}

\section{Concluding Remarks}

We have derived secular evolution equations, Eqs. (\ref{angledotave}) and (%
\ref{psipdotave}) for the angular variables $\alpha,\phi_n$ and $\psi_p$
characterizing the orientation of the orbit, with the inclusion of the
leading order relativistic and spin-orbit coupling contributions. Together
with the evolution of the angles $\kappa_{i}$ given by the first two Eqs.
(2.17) of \cite{GPV3} and the projections of $\mathbf{J=L+S}_{\mathbf{1}}+%
\mathbf{S}_{\mathbf{2}}$, determining $\psi_i$ in terms of the other
variables, these form a closed system of differential equations for the
conservative evolution of the angular variables.

We have complemented these with the equations expressing the respective
secular losses of the linear momentum components, Eqs. (\ref{Pdotave}). Then
we have a closed system of coupled first order differential equations,
giving the linear momentum loss during the inspiral in analytic~form.

This system of equations is also suitable for numerical evolution, which in
principle can lead to the value of the recoil velocity right before the
plunge and also allows to determine the shift in the position of the binary
during the inspiral. Our analytic approach is suitable for discussing the
dependence of the recoil on various parameters characterizing the binary.

In order to gain higher accuracy, the results derived in this paper can be
further generalized by the inclusion of higher order corrections: the
spin-spin, mass quadrupole - mass monopole, and second order relativistic
contributions.

\section*{Acknowledgements}

This work was supported by the Hungarian Scientific Research Fund (OTKA)
grants no. 69036 and 68228, also by the Pol\'{a}nyi and Sun Programs of the
Hungarian National Office for Research and Technology (NKTH). L\'{A}G is
grateful to the organizers of the Amaldi8 meeting for support.

\end{document}